# Magnetic properties of Co/Cu/Py antidot films

J. C. Denardin[1,2], E. Burgos[3], R. Lavín[4], S. Vojkovic[4], J. Briones[1,2*] and M. Flores[5]

[1]Departamento de Física, Universidad de Santiago, Av. Ecuador 3493, Estación Central, Santiago, Chile

[2]Center for the Development of Nanoscience and Nanotechnology, CEDENNA, 9170124, Santiago, Chile

[3]Facultad de Ciencias, Universidad de Chile, Las Palmeras 3425, Ñuñoa, Santiago, Chile.

[4]Instituto de Ciencias Básicas, Universidad Diego Portales, Av. Ejército 441, Santiago, Chile

[5]Departamento de Física, FCFM, Universidad de Chile, Blanco Encalada 2008, Santiago, Chile.

The magnetic properties of ordered nanoscale Co/Cu/Py multilayer antidot films with different pore sizes prepared on top of nanoporous alumina membranes are presented. By using Co and Py films separated by a Cu thin layer, and tuning the pore diameters of the NAMs we were able to play with the coercivity of the films and observe stepped magnetization curves, as a consequence of the different coercivities of the Co and Py films. The magnetic properties of the multilayer antidots have been measured and compared with results obtained for antidots of Cu/Py. The magnetization reversal process that occurs in each individual layer and in the multilayer was studied by means of micromagnetic simulations.

---

[*] Email: joel.briones@usach.cl

I.  INTRODUCTION

Nanostructured magnetic elements have received much attention from the scientific community in the last two decades due to their potential applications, ranging from sensors for the electronic and electromechanical industry to the storage media for the magnetic recording industry. The magnetic nanostructures usually referred as antidots are based on magnetic thin films with periodic arrays of holes. In the last years the study of magnetic nanoparticles based on arrays of antidots have attracted intensive attention because they are a promising candidates for a new generation of ultra-high-density magnetic storage media, and an exciting topic in fundamental physics [1-33]. In an antidot array, magnetic features such as coercive field, anisotropy axes and reversal mechanisms, among others, can be tailored by tuning the geometric parameters of the array [1-4]. Generally, the studies have been focused in antidot systems of single layered magnetic materials [1-24]; and more recently on multilayered ones [25-33]. Antidots of NiFe (Py) [1,5-16] and Ni [17,18] films are widely studied presenting an increased coercivity compared to the continuous form. Antidots made of Co present a large enhancement of the coercivity [2-4,19-24], and are usually deposited with Pt for out of plane recording purposes [25-27]. Arrays of Co/Cu/NiFe, CoFe/Cu/NiFe and [Co/Cu]$_N$ multilayered antidots have been studied previously by means of magneto-transport measurements [28-30]. In this work we present the magnetic properties of ordered nanoscale Co/Cu/Py antidots arrays with different pore sizes prepared on top of nanoporous alumina membranes (NAMs). By using Co and Py films separated by a Cu thin layer, and tuning the pore diameters of the NAMs we were able to play with the coercivity of the films and observe stepped magnetization curves, as a consequence of the different coercivities of the Co and Py films.  By systematic measurements of the magnetic and magnetotransport properties of these samples, and comparing with simulations, we were able to understand in more detail the magnetization reversal that takes place in these multilayer antidots.

## II. EXPERIMENTAL PROCEDURE

Nanoporous alumina membranes were prepared from a 0.32 mm thickness aluminum foil (Good-Fellow, 99.999 %) by using the so-called two-step anodization technique [34]. The pore diameter was modified by a posterior treatment in a 5W% $H_3PO_4$ solution at 35 °C for 5-10 minutes producing samples with 40, 50, 65 and 75 nm of pores diameter. The films were deposited by sputtering on top of NAMs that have been previously anodized in oxalic acid. The thickness of the films was controlled by a crystal balance, and all the samples where sputtered at the same time, having 10 nm of Co, 5 nm of Cu and 20 nm of Py (NiFe). Another layer with 8 nm of Cu was deposited on top of the films to avoid oxidation. The magnetic measurements where performed in a homemade Alternating Gradient Magnetometer (AGM) at room temperature. Magnetoresistance of the samples was measured in a Cryogenic LTD mini 5T VSM, using the Resistivity probe (RnX) measuring platform.

Micromagnetic simulations were made using the 3D OOMMF package [35]. The magnetization of the sample is computed as a function of time using damped Landau-Lifshitz-Gilbert equation of motion (LLG) of the magnetization at zero temperature [36]. Our starting point is a square film of area of 1μm x 1μm and a separation between holes of 100 nm. Thus, the ferromagnetic system is spatially divided into cubic cells of 2 nm x 2 nm x 2 nm (smaller than the exchange lengths of materials), and within each cell the magnetization is assumed to be uniform. The micromagnetic simulations are performed using typical Co and Py parameters: saturation magnetization $M_{S,Co}$ = 1.4 x $10^6$ A/m and $M_{S,Py}$ = 8.6 x $10^5$ A/m, exchange stiffness constant $A_{Co}$

= 3 x $10^{-11}$ J/m and $A_{Py}$ = 1.3 x $10^{-11}$ J/m, and anisotropy constant of K = 0 (for polycrystalline films). In all the cases the damping parameter was chosen as 0.5.

III. RESULTS AND DISCUSSION

Fig. 1 shows the SEM images of the Co/Cu/Py films deposited on top of the NAMs. The magnetic films follow the morphology of the patterned membrane, and the area between pores decreases as the pore sizes increases, changing the space available for nucleation and propagation of the domain walls. By using Co and Py films separated by a Cu thin layer, and tuning the pore diameters of the NAMs we were able to play with the coercivity of the films and observe stepped magnetization curves, as can be observed in Fig. 2. It was observed that the magnetization reversal of the Co and Py layers occurs at different fields, and the difference in coercivities increases with pore diameter. On the same set of NAMs we also deposited films of Cu (15 nm)/Py (20 nm). In these films we observe a small increase in the coercivity with increase of pore diameter, but do not observe the two-step magnetization reversal as in the Co/Cu/Py films (see Figure 2-b). In the continuous Co/Cu/Py films, deposited on glass, there is a small two-step effect in the curves, but very small coercivities are observed. It is worth to note that the magnetization reversal of the Co/Cu/Py multilayer antidots occurs at fields similar to the Cu/Py antidots (their coercivity are comparable). The second step in the hysteresis curve, which appears at a higher field, is the magnetization reversion of the Co layer. Antidots of Co would then have a higher coercivity, which is reduced by the interaction with the Py magnetic layer. Since the separation among de Co and Py layers is of only 5 nm, magnetostatic and RKKY interactions could be present [37], resulting in a decrease in the coercivity as observed in the experimental curves.

In order to understand in more detail the magnetization reversal in these multilayers we have also studied the magnetization reversal process by means of micromagnetic simulations using OOMMF [35]. In order to perform the micromagnetic simulations, SEM images as from Figure 1 were treated by image filters and used as masks in OOMMF [9,24], allowing us reproducing the typical defects observed in the samples.

The thicknesses of the layers in the simulation are 12 nm of Co and 20 nm of Py, separated by a gap of 4 nm. Figure 3 (a) shows the simulated hysteresis curves for the Co/gap/Py antidots film (gap is of the same thickness of the Cu layer in the experimental film) in function of the pore diameter, and the inset of Figure 3 shows the coercivity in function of the pore diameter for the Co/gap/Py antidots film, and antidots isolated films of the Co and Py. The coercivity values of the simulated curves are higher than the observed in the experimental ones, and the steps in the magnetization reversal are not so pronounced as in the experimental curves. Despite the difficulties in reproduce in the simulations all the aspects and complexities of the experimental samples, the results can be useful to observe and understand the domain wall propagation in the different layers.

Figure 4 shows the simulated domain configuration in the demagnetized state for different pore diameters for the Co/Gap/Py antidots films. These images are snapshots of the magnetic configuration at –Hc (negative coercivity) state, obtained when the external magnetic field is decreased from +Hs (saturation), and starting with the antidot films saturated in the x direction. When the size of the pores increase, and the space for the propagation of the domain walls is reduced, the interaction among different propagating domains produces multiple smaller domains and consequently gives rise to a more complex magnetic domain structure, which results in an increase in the coercivity of the films. It is also possible to observe in the images of

Figure 4 the presence of some kind of vortex behavior in the region between some pores. Vortex switching could also be responsible for the reduction of coercivity in the multilayers. In the Co layer the interactions between different domain walls is stronger than in the Py layer, and the magnetization reversal of the Co layer will occur at larger external fields.

For Co/Gap/Py simulated antidot systems we obtain a lower coercivity than for antidot systems of a single magnetic material (see Fig. 3 (inset)). This is because the Py antidot film inverts its magnetization at a lower external field than the Co antidots films, and induces the Co film to inverts its magnetization at lower external field, due to the magnetic coupling between both layers. This magnetic coupling is clearly observed in Figure 4, in the similarity of the magnetic patterns for the Co and Py layers. The magnetostatic interaction favors an antiferromagnetic coupling among the Co and Py layers, being responsible for a loop bifurcation that is also observed in the simulated curves.

Finally, magnetoresistance (MR) measurements have been performed in the samples. Figure 5 shows the MR curve of the Co/Cu/Py antidot sample with 75 nm pore diameter, measured with the magnetic field transversal to the current (TMR) configuration. In this graph the magnetization curve of the same sample is also shown for comparison.

The MR curve shown in Figure 5 follows the magnetization reversal and does not show the complex peak structure observed by other authors in similar antidot multilayers [28]. The MR response of the sample with the field parallel to the current (longitudinal MR, or LMR) is similar to the one shown in Figure 5 and does not show anisotropic MR (AMR) effect.

The effect of interlayer exchange coupling in MR curves has been studied previously in Co/Cu/Co continuous films [30], and complex MR curves have been observed when the Cu thickness was 5 nm, corresponding to an antiferromagnetic coupling among the layers. In antidots fabricated by anodization process, the roughness of the AAO topography is of the order of 10 nm [38], and plays an important role in the magnetotransport properties. It is known that the interlayer exchange coupling strongly dependent of the interlayer roughness, being the roughness responsible for the appearance of dipolar interaction between the layers [37]. The isotropic MR response that is observed in these AAO antidots could be consequence of the topography of the samples.

IV. CONCLUSION

By systematic measurements of the magnetic properties of Co/Cu/Py and Cu/Py multilayer antidots, and comparing the results with simulations, we were able to understand in more detail the magnetization reversal that takes place in these systems. The Co/Cu/Py antidots present hysteresis curves with two steps that correspond to the magnetization reversal of the Py and Co layers. The progressive increase in the distribution of coercivities and interactions gives rise to more complex magnetic domain structures when the diameter of the pores increase, which results in an increasing in the coercivity of the films. On the other hand, the interaction between different magnetic layers decreases the coercivity of the multilayer films.

ACKNOWLEDGMENTS

The support from Fondecyt Grants 3120059, 1140195 and 11110130; Millennium Science Nucleus, Basic and Applied Magnetism Grant N°P10-061-F, and CONICYT BASAL CEDENNA FB0807, is gratefully acknowledged.

Figures Captions.

Fig. 1. Co/Cu/Py multilayer films on antidots with pores of different diameters a) 40 nm, b) 50 nm, c) 65 nm and d) 75 nm.

Fig. 2. Magnetization curves of Cu/Py (up) and Co/Cu/Py (down) multilayer films on antidots with pores of different diameters.

Fig. 3. Simulated magnetization curves for antidots of Co/Cu/Py with different pore diameters and coercivity of the Co, Py and Co/Cu/Py antidots as function of pore diameter (inset).

Fig. 4. Simulated images of the magnetization state of Co and Py antidot films with different pore diameters.

Fig. 5. Magnetoresistance curve of the Co/Cu/Py antidot sample with 75 nm pore diameter (filled red symbols). The magnetization curve of the same sample is also shown for comparison (open green symbols).

Figures:

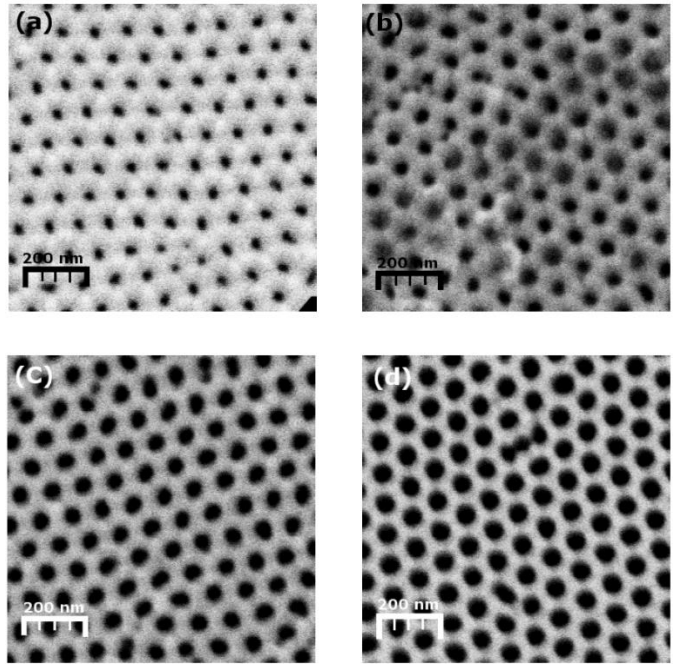

Fig. 1

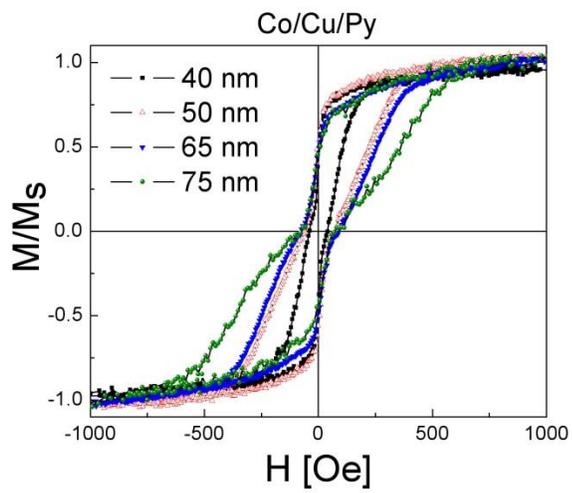

Fig. 2

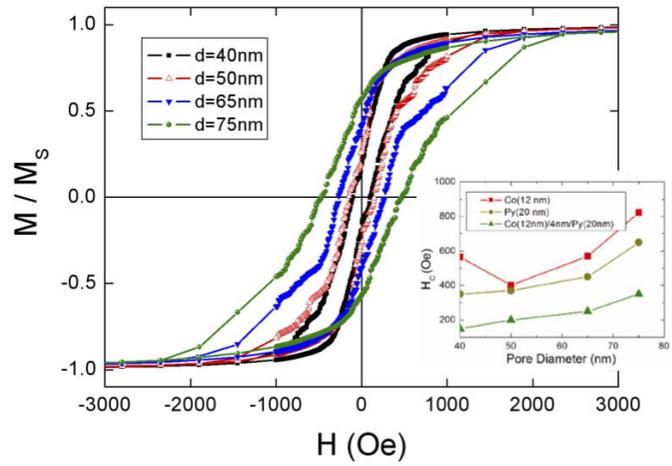

Fig. 3

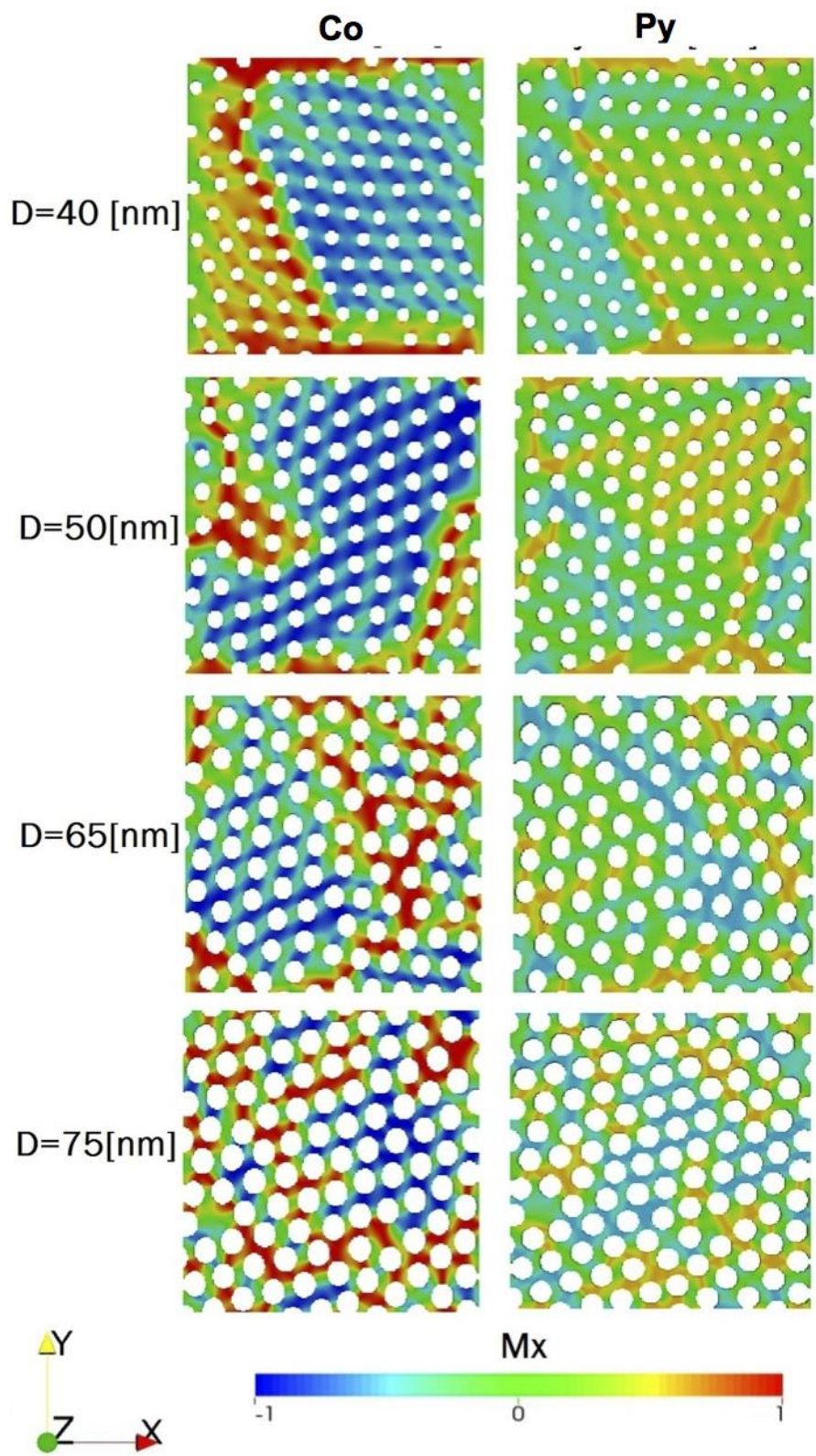

Fig. 4

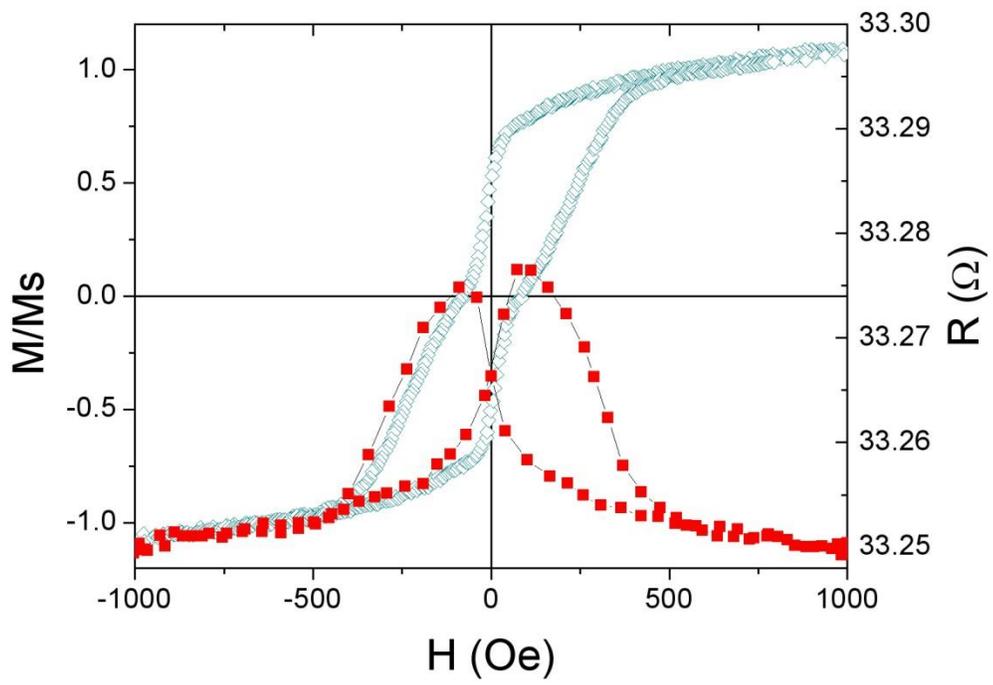

Fig. 5